\documentstyle[twocolumn,prl,aps]{revtex}

\newcommand{\be}{\begin{equation}}
\newcommand{\ee}{\end{equation}}
\newcommand{\ba}{\begin{eqnarray}}
\newcommand{\ea}{\end{eqnarray}}

\title{Comment on ``Electron Mass Operator in a Strong Magnetic Field and
Dynamical Chiral Symmetry Breaking"} 

\author{V.P.~Gusynin}

\address{Institute of Theoretical Physics
Sidlerstrasse 5, CH-3012 Bern, Switzerland\\
and Bogolyubov Institute for
Theoretical Physics, 252143, Kiev, Ukraine}

\author{V.A.~Miransky}

\address{Department of Applied Mathematics, University of Western
Ontario, London, Ontario N6A 5B7, Canada}

\author{I.A.~Shovkovy}

\address{School of Physics and Astronomy, University of Minnesota,
Minneapolis, MN 55455, USA}

\date{\today}
\draft
\begin{document}

\maketitle


\begin{abstract}
This is a comment on the paper ``Electron Mass Operator in a Strong Magnetic 
Field and Dynamical Chiral Symmetry Breaking" by A. V. Kuznetsov and 
N. V. Mikheev [Phys. Rev. Lett. 89 (2002) 011601]. We show that the main
conclusions of the paper are incorrect. 
\end{abstract}

Magnetic catalysis of chiral symmetry breaking \cite{prl73} is a well
established phenomenon in $(2+1)$ and $(3+1)$ dimensional relativistic
models. The general result is that a constant magnetic field leads to the
generation of a fermion dynamical mass even at the weakest attractive
interaction between fermions. The essence of the effect is the dimensional
reduction $D\to D-2$ in the dynamics of fermion pairing in a magnetic
field \cite{prl73}.

The realization of this phenomenon in quantum electrodynamics (QED) was
studied in detail in
Refs.~\cite{prd4747,Leung,Hong,Smilga,prl83,npb563,Alex}. In particular,
in Refs.~\cite{prl83,npb563}, we derived an asymptotic expression for the
fermion dynamical mass in the chiral limit in QED, reliable for a weak
coupling.  In a recent Letter \cite{KM}, Kuznetsov and Mikheev attempted
to revise that analysis. These authors claimed that:

\begin{enumerate} 

\item[(i)] In a model with $N$ charged fermions, a critical number $N_{cr}$
exists for any value of the electromagnetic coupling constant, such that
the dynamical mass does not arise for $N>N_{cr}$, and thus the chiral
symmetry remains unbroken.

\item[(ii)] The fermion dynamical mass is generated with a double splitting for
$N<N_{cr}$.

\end{enumerate}

In this Comment, we will show that these conclusions of Ref.~\cite{KM}
are incorrect. In fact, we show that

\begin{enumerate}

\item[(i$^{\prime}$)] The generation of the fermion dynamical mass takes
place for any number of the fermion flavors $N$ if the magnetic field is
less than the value of the Landau pole, i.e., if the theory itself is well
defined.  The erroneous conclusion (i) of Ref.~\cite{KM} follows from an
inappropriate treatment of the Schwinger-Dyson (gap) equation for large
$N$.

\item[(ii$^{\prime}$)] There is a unique solution for the dynamical fermion
mass in any given theory with a fixed value of $N$.

\end{enumerate}

Let us start by mentioning that the gap equation (12) of Ref.~\cite{KM} 
for the case of zero bare mass of fermions,
\be
\frac{\alpha_R}{2\pi} 
\left(\ln\frac{\pi}{N \alpha_R} -\gamma_{E}\right)
\ln\frac{|eB|}{m^2} =1,
\label{gap-eq}
\ee
is the simplest approximation to the gap equation previously derived in our
papers \cite{prl83,npb563} in the limit of a weak coupling and not too
large values of $N$. Here, by definition, $m\equiv m_{dyn}(|eB|)$ is the
dynamical mass and $\alpha_{R}$ is the coupling constant at the scale of
$\sqrt{|eB|}$.  One should note that the validity of the above equation 
breaks down when the number of fermions becomes large.  Indeed, with the
increasing value of $N$ the photon mass squared $M_{\gamma}^{2}
=2N\alpha_{R}|eB|/\pi$ should eventually 
become comparable to the magnetic field scale 
$|eB|$. When this happens the leading-log approximation used
in Ref.~\cite{KM} fails, and therefore one cannot trust equation
(\ref{gap-eq})  any longer. In fact, this could be seen from
Eq.~(\ref{gap-eq}) itself: a simple-minded large $N$ limit leads to the
change of sign on the left hand side of the equation, suggesting that no
reasonable solution for the dynamical mass exists. Obviously, this is
incorrect because the original integral form of the gap equation [see
Eq.~(54) in Ref.~\cite{npb563}] does not share this unphysical property.  
It is not difficult to derive an approximate algebraic form of the gap
equation which works reasonably well for all values of $N$. 
Its explicit form reads
\be
-\frac{\alpha_R}{2\pi} 
\exp\left(\frac{N\alpha_R}{\pi}\right) 
\mbox{Ei}\left(-\frac{N\alpha_R}{\pi}\right)  
\ln\frac{|eB|}{m^2} =1,
\label{cor-gap-eq}
\ee
where $\mbox{Ei}(z)$ is the exponential integral function. Notice, that
this gap equation has a nontrivial solution for any number of fermions. In
particular, in the limit of large $N$, the corresponding solution for the
dynamical mass reads
\be
m \simeq \sqrt{|eB|} \exp\left(-N\right), \quad
\mbox{for} \quad N\gg \pi/\alpha_{R}.
\ee
This demonstrates that the conclusion of Ref.~\cite{KM} about the 
existence of a critical value of $N$ is incorrect. 

For clarity, let us emphasize that the only ``assumption'' in the above
derivation was that the scale of the magnetic field lies below the scale
of the Landau pole (mathematically, this is expressed as
$\alpha_{R}<\infty$). This condition is of course necessary because the
theory is not well defined otherwise.

Another claim of Ref.~\cite{KM} is that ``a fermion dynamical mass is
generated with a doublet splitting for $N < N_{cr}$". In fact, this 
conclusion was reached as a result of misinterpretation of the 
following feature of the gap equation (\ref{gap-eq}): this equation has 
two solutions with different values of the fermion mass for each 
choice of $|eB|$, $N$ and the coupling constant $\alpha \equiv 
\alpha(m)$ related to the renormalization group (RG) scale $\mu=m$. 
It is important that $m$ is the {\it dynamical} mass, determined 
from the gap equation, and that in Ref. \cite{KM}
the running coupling is taken from the one-loop RG 
equations in QED {\it without} a magnetic field.  
Let us show 
that the interpretation of this feature as a mass splitting is 
incorrect. To see this, we turn to the RG equation for the running 
coupling constant:
\be
\alpha(m) = \frac{\alpha(\mu)}{1-\alpha(\mu) b_{0}
\ln(m^2/\mu^2)}, \quad \mbox{with} \quad
b_{0} =\frac{N}{3\pi}.
\label{RG}
\ee
Now, suppose that indeed, for given values of $|eB|$ and $N$, 
the gap equation can have two solutions with different masses, 
$m_1$ and $m_2$, and that the constraint $\alpha(m_1)=\alpha(m_2)$ 
can be satisfied at the same time. Then, it is easy to see 
that these two solutions correspond not to the same theory, as the authors 
of Ref.~\cite{KM} believe, but to two different theories. Indeed, since 
$\alpha(m_1) = \alpha(m_2) \equiv \alpha$, the values of the two 
corresponding RG coupling constants $\alpha_1(\mu)$ and $\alpha_2(\mu)$ 
related to the {\it same} scale $\mu$ are different. The latter directly 
follows from equation (\ref{RG}):
\be
\alpha_{i}(\mu) = \frac{\alpha}{1+\alpha \; b_{0}
\ln(m_{i}^2/\mu^2)}, \quad
i = 1, 2.
\ee
Therefore, the solutions $m_{1}$ and $m_{2}$ correspond to two theories 
with different RG coupling constants. In those theories, in particular, 
the coupling constants take different values at the scale $\mu=\sqrt{|eB|}$.
This in turn implies that there is only one solution for the dynamical 
fermion mass in any given theory with a fixed value of $N$.

We emphasize that there is nothing wrong with the parameterization of a 
solution in QED in a magnetic field by a coupling constant in QED without
the magnetic field. One can use a coupling constant related to any fixed
scale $\mu$. The choice of the scale $\mu = m$ in Ref.~\cite{KM}, although
possible, is contrived: this scale can be determined self-consistently only
after solving the gap equation, and therefore it is not a free parameter. 
This subtlety led the authors to the incorrect interpretation regarding the
appearance of the ``mass splitting" \cite{footnote1}.

In connection with this, we would like to point out that the claim in 
Ref.~\cite{KM} that the coupling constant renormalization was not 
taken into account in Refs.~\cite{prl83,npb563} is without basis. 
Along with the polarization effects, the renormalization of the coupling 
constant was taken properly into account in Refs.~\cite{prl83,npb563}.
In fact, the choice of the coupling constant $\alpha_{R}$ related to
the RG scale $\mu=\sqrt{|eB|}$ in Refs.~\cite{prl83,npb563} is of 
course nothing else but a coupling constant renormalization. This 
choice is the most convenient (although not unique) parameterization 
of the gap equation and its solution. Indeed, $\sqrt{|eB|}$ is the only 
free dimensional parameter in this problem \cite{footnote}.

In conclusion, the magnetic catalysis of chiral symmetry breaking is
always realized as soon as the theory itself is well defined. In the 
case of QED, this simply requires that the values of magnetic fields
should lie below the scale of the Landau pole.

{\bf Acknowledgments.} The research of V.P.G. has been supported in part
by the National Science Foundation under Grant No. PHY-0070986 and by the
SCOPES-projects 7~IP~062607 and 7UKPJ062150.00/1 of Swiss NSF. V.A.M. is
grateful for support from the Natural Sciences and Engineering Research
Council of Canada. The work of I.A.S. was supported by the U.S. Department
of Energy Grant No.~DE-FG02-87ER40328.


\begin{references}
\bibitem{footnote1} It is appropriate to mention that even the 
existence of two solutions with $\alpha(m_{1})=\alpha(m_{2})$
is no longer valid after the corrected gap equation (\ref{cor-gap-eq}) 
is considered.
\bibitem{footnote} Another advantage of this choice is that the
value of the coupling related to the scale $\sqrt{|eB|}$ 
(as well as to any scale above it) is essentially the same in 
QED with and without the magnetic field \cite{npb563}.
\end{references}
\end{document}